\documentclass[showpacs,twocolumn,prb,aps]{revtex4}
\usepackage{epsfig}
\usepackage{amsmath}
\usepackage{amssymb}
\begin{document}
\title{Stochastic Cluster Series expansion
 for quantum spin systems}

\author{Kim Louis and C. Gros} 

\affiliation{Fakult\"at 7, Theoretische Physik,
 University of the Saarland,
66041 Saarbr\"ucken, Germany.}

\date{\today}

\begin{abstract}
In this paper we develop a cluster-variant of the Stochastic
Series expansion method (SCSE). For certain systems with longer-range
interactions the SCSE is considerably more efficient than
the standard implementation of the Stochastic Series Expansion (SSE),
at low temperatures. As an application of this method we calculated 
the $T=0$-conductance for a linear chain with  a (diagonal) next nearest neighbor 
interaction.

\end{abstract}
\maketitle

{\em Introduction}---
The development of efficient Quantum Monte Carlo (QMC) methods
with loop-updates, like the original loop-algorithm \cite{Everts} 
and the SSE \cite{Sandvik,Sandvik2,Dorneich,SandSyl},
has been a major advancement. They are very efficient for
the anistropic Heisenberg models, like the $xxz$-chain and
can be generalized to more complicated Hamiltonians, but in some cases 
only with reduced performance.

Here we study the $xxz$-chain with a 
diagonal next nearest neighbor interaction.
This model is better suited for the description of fermionic systems
since it takes into account that the interaction (Coulomb-repulsion)
is long ranged. Furthermore, it is one of the simplest non-integrable systems.
Consequently, this model has attracted the attention of many authors.
\cite{zopr,andrei,Kats} 

We find that a standard SSE-implementation performs only poorly on our
model system. The reason for this lies in the fact that a certain
transition in the operator loop update---called ``bounce''---
is given a relatively large weight.
Sylju{\aa}sen and Sandvik have shown that such a large bounce
may affect the efficiency of the algorithm---especially at low
temperatures. Hence, one should strive to find a way to reduce 
the bounce weight.

Here we report,
that a new implementation of the SSE-algorithm, the cluster-SSE, yields a 
considerably smaller bounce weight than the standard SSE, improving
thus the efficiency considerably at low temperatures.

\begin{figure}
\epsfig{file=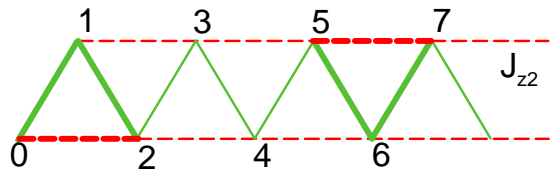,width=0.4\textwidth,angle=0}
\caption{ The model Hamiltonian that will be discussed in this
 paper. Solid lines indicate a full Heisenberg-like
 interaction between the sites; dashed  lines stand for sites coupled 
only by a $z$-$z$-term (Ising-like interaction).
Two  clusters---as defined in the text--- are indicated by thicker lines. 
}
\label{hsplit}\end{figure}

{\em Conventional SSE}---
We will now briefly discuss the conventional SSE-implementation
 and point out its difficulties.
Our model of interest is a frustrated chain with a diagonal
next-nearest neighbor interaction: (see  Fig.\ \ref{hsplit})
\begin{eqnarray}\nonumber
H&=&\sum_n\Big\{\frac{J_x}{2}(S_n^+S_{n+1}^-+S_n^-S_{n+1}^+) \\
 &&\ \ \ +\ J_zS_n^zS_{n+1}^z +J_{z2}S_n^zS_{n+2}^z\Big\}.
\label{hamilfrus}
\end{eqnarray}
Note that only the interaction part is frustrated such that
a Monte Carlo simulation  will not suffer from the sign 
problem.
Following Ref.\ \onlinecite{Sandvik}
we start by splitting the Hamiltonian into a set of local
operators, i.e., $H=\sum_{h\in\frak{h}}h$ with
$\frak{h}=\{h_{n,n+1}^{(t)},h_{n,n+2}^{(4)}:t=1,2,3;n\in{\mathbb N}\}$ 
and
\begin{equation}
\begin{array}{rcl}
h_{n,n+1}^{(1)}&=&J_xS_n^+S_{n+1}^-/2,\\
h_{n,n+1}^{(2)}&=&J_xS_n^-S_{n+1}^+/2,\\
h_{n,n+1}^{(3)}&=&C+J_zS_n^zS_{n+1}^z,\\
h_{n,n+2}^{(4)}&=&C_2+J_{z2}S_n^zS_{n+2}^z,
\end{array}
\label{h_nn}
\end{equation}
where the constants $C$ and $C_2$ are needed to ensure that all
 matrix elements between $S^z$-eigenbasis-states are positive.
Using the Taylor expansion, the partition function may be written as
\begin{equation}Z=\sum_{M,\phi,\alpha}\frac{(-\beta)^M}{M!}
\prod_{m=1}^M\,
\langle \alpha_m,\phi(m)\alpha_{m+1}\rangle
\label{part}
\end{equation}
where $M\in\mathbb N$, $\phi:\{1,\dots,M\}\to \frak{h}$, and
the $\alpha_m$
run over all $S^z$-eigenbasis states with periodic
boundary-conditions
$|\alpha_1\rangle=|\alpha_{M+1}\rangle$
along the imaginary-time axis.
The factors 
$\langle \alpha_m,\phi(m)\alpha_{m+1}\rangle$
in Eq.\ (\ref{part}) are called {\em plaquettes} and
are non-negative due to (\ref{h_nn}).
Hence, we can obtain the partition function by  sampling
the terms in Eq.\ (\ref{part}) with their relative weight factors
over all spin-configurations $|\alpha_m\rangle$ and
local operators $\phi(m)\in\frak{h}$.

Since each operator $\phi(m)$ acts only on two sites
we may write 
$$W_{n_1,n_2}(s_{n_1}^m,s_{n_2}^m,s_{n_1}^{m+1},s_{n_2}^{m+1}):=\langle\alpha_m,
\phi(m)\alpha_{m+1}\rangle$$
where   $s_{n}^m=S_n^z|\alpha_m\rangle$ are spin variables and $W_{n_1,n_2}$
is given by the prefactors of (\ref{h_nn}).
Note that because of translation invariance $W_{n_1,n_2}$ depends
actually only on $n_2-n_1$.
For brevity we will in the sequel denote
the four arguments, which we call plaquette {\em legs},
 by one super-argument,
 which we call the plaquette {\em state}.
The SSE knows two updates:

\begin{itemize}
\item In the {\em diagonal update}
an insertion/removal of one plaquette is considered.

\item The {\em loop update} (constructs and) flips  a sub-set of the
spin variables such that the new configuration is allowed--- that means
      has non-zero weight.
\end{itemize}

The loop update may be achieved step by step.\cite{Sandvik} 
In each step we consider only one plaquette, 
whose state will be changed from $i$ to $j$ by flipping
two legs $l_i$ and $l_j$.
The ``in-going'' leg $l_i$ is given
and we have to choose the ``out-going'' leg $l_j$
with a certain 
probability   $p(i\to j,l_i,l_j)$. 
Of course, the probability for a
transition $i\to j$ must be zero, if one
of the states $i$ or $j$ has zero weight.

These probabilities---yet to be determined---will play a central r\^ole in
what follows. Therefore, we will explain in detail how they are
obtained.

They may  equivalently be represented by
$(2^4\cdot 4)\times(2^4\cdot 4 )$-matrices $a^{n_1,n_2}$ with entries  
$$a^{n_1,n_2}_{(i,l_i)(j,l_j)}:=p(i\to j,l_i,l_j)W_{n_1,n_2}(i).$$ 
The fact that the  $p(i\to j,l_i,l_j)$ are probabilities
imposes  for each $l_i$ and $i$ the following
conditions: 
\begin{equation} W_{n_1,n_2}(i)=\sum_{j,l_j}a^{n_1,n_2}_{(i,l_i)(j,l_j)}.
\label{vectoreq}\end{equation} 
The detailed balance condition is satisfied if the matrices $a^{n_1,n_2}$
are  symmetric. 

Closer inspection shows that the matrices $a^{n_1,n_2}$ are block
diagonal where the dimension of the blocks $d$
equals half the number of allowed
plaquette states. Hence,
if $n_2-n_1=1$, we have $d=3$ (as for the 
$xxz$-chain\cite{SandSyl};
it would be four if pair creation and
pair annihilation were not forbidden\cite{Dorneich})
whereas $d=2$ for $n_2-n_1=2$.

All the elements in one single block  of $a^{n_1,n_2}$ differ in at least 
one state index. Hence, we may drop the leg indices $l_i$ and $l_j$
without causing ambiguities. Moreover, we will also omit the explicit
indication of the site
indices $n_1$ and $n_2$ in $a$ and $W$.

For the blocks of length $3$ a formal solution for the non-diagonal entries
 $a_{ij}$ which meets detailed balance and Eqs.\ (\ref{vectoreq})
 can easily be stated\cite{SandSyl}
\begin{equation}
a_{ij}=[W(i)+W(j)-W(k)]/2+[a_{kk}-a_{ii}-a_{jj}]/2
\label{soldim3}\end{equation}
where $k$ is the third plaquette state in the same block of $a$ as $i$
 and $j$. 
One sees that the diagonal entries of the matrices $a$ 
remain free---apart from the final restriction that
 all probabilities (entries
 of $a$) need be positive.

Ref.\ \onlinecite{SandSyl} tells us
how to dispose of the remaining degrees of freedom:
The diagonal entries of $a$ 
(baptized bounces\cite{SandSyl} since they correspond to 
the choice $l_j=l_i$)
are always allowed, but  turn out to be inefficient
as they impede the update scheme.
As a general rule,\cite{SandSyl} one can say that the most favorable 
among all possible solutions to Eqs.\ (\ref{vectoreq}) is the one 
with minimal diagonal entries.

In  the blocks of length three all bounce weights may be put to
zero for a wide parameter range,\cite{SandSyl}
but in the blocks of dimension two
Eqs.\ (\ref{vectoreq}) dictate one of the two bounces $a_{ii}$ and
$a_{jj}$ to be equal to $|W(i)-W(j)|=J_{z2}/2$. A situation which is far from optimal.

\begin{figure}
\epsfig{file=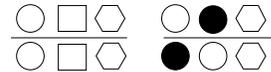,width=0.2\textwidth,angle=0}
\caption{ Pictorial representation of the plaquettes similar to the one
 used in Ref.\ \onlinecite{SandSyl}. The operator itself is indicated by
 the horizontal central bar. The symbols stand for the six legs of the
 plaquettes. Equal symbols mean equal spins, and filled and empty
 symbols mean opposite spins. In this way the left diagram 
 represents the eight diagonal plaquettes, and the right diagram, four
 non-diagonal plaquettes. The remaining non-diagonal plaquettes may be
 obtained by reflection with respect to the vertical central
 plaquette-axis
 and are
 not shown here.
}
\label{allplaqs}\end{figure}

{\em Cluster variant (SCSE)}--- The problem with the SSE method
is that some plaquettes---associated with operators  $h^{(4)}$---
have only
four allowed (diagonal) states.
 This can be avoided by splitting
the Hamiltonian into small clusters instead of mere two-sites operators.
We now introduce the Stochastic {\em Cluster} Series expansion (SCSE).
In the case of the frustrated chain we split the Hamiltonian (\ref{hamilfrus}) not into two-
sites but three-sites operators (see Fig.\ \ref{hsplit}):
\[
\begin{array}{rcl}
h_{n,n+1,n+2}^{(1)}&=&J_xS_n^+S_{n+1}^-/4\\
h_{n,n+1,n+2}^{(2)}&=&J_xS_n^-S_{n+1}^+/4\\
h_{n,n+1,n+2}^{(3)}&=&J_xS_{n+1}^+S_{n+2}^-/4\\
h_{n,n+1,n+2}^{(4)}&=&J_xS_{n+1}^-S_{n+2}^+/4\\
h_{n,n+1,n+2}^{(5)}&=&J_z/2(S_n^zS_{n+1}^z+S_{n+1}^zS_{n+2}^z)+\\
&&\ +J_{z2}S_n^zS_{n+2}^z~+ C.
\end{array}
\]
Note the factor $1/4$ instead of $1/2$ for $h^{(t)}$ with $t=1,2,3,4$.
It stems from the fact that each non-diagonal operator is ``distributed''
between two clusters. The possible plaquette states are shown in 
Fig.\ \ref{allplaqs}.

The diagonal update and loop construction remain unchanged 
with respect to the SSE.  In the loop update the only difference 
is that $W$ is now a function of $6$ variables. 
Let us now turn to the matrix $a$. The transitions of the form
$h^{(t)}\to h^{(r)}$, when $t\in\{1,2\}$ and
$r\in\{3,4\}$ will be ruled out from the beginning (for simplicity and because we do not
expect that they will give  assistance in minimizing the bounce.)
All other transitions may have positive probability.

\begin{figure}
\epsfig{file=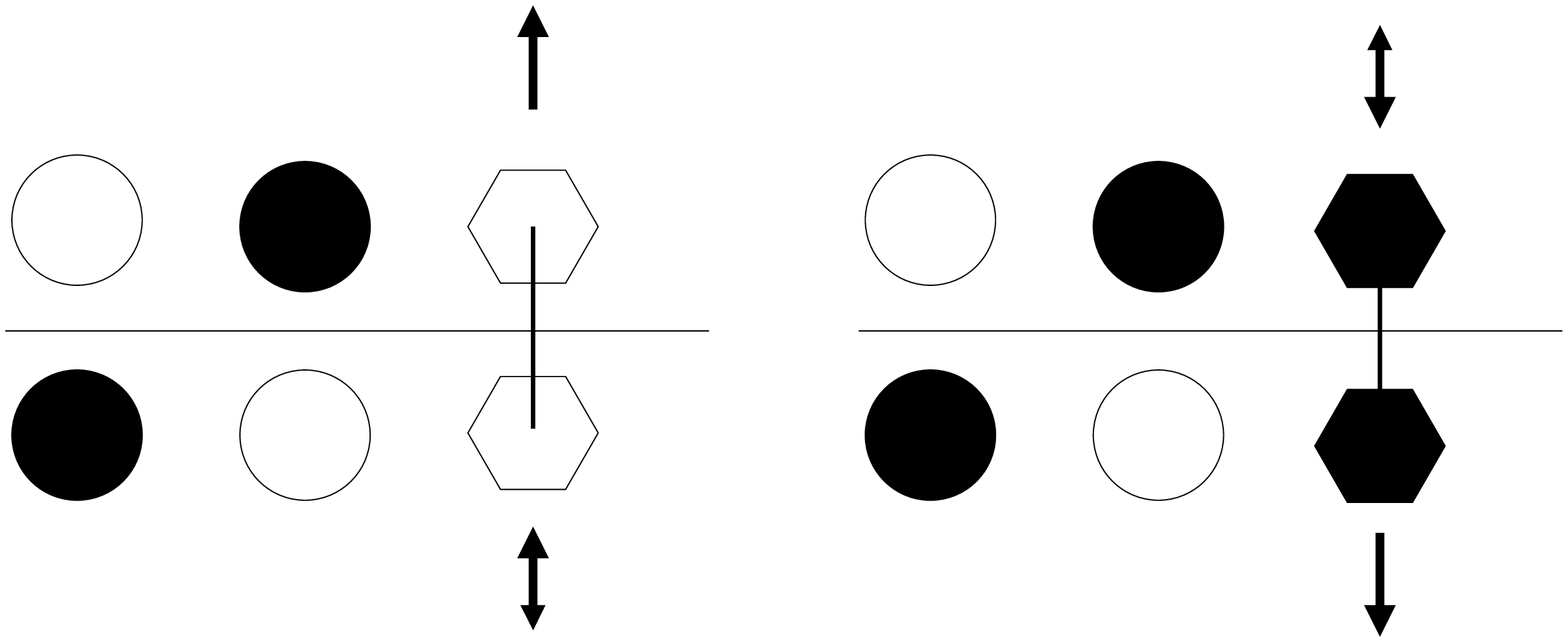,width=0.2\textwidth,angle=0}
\rule{\hsize}{2pt}

\vspace{0.2cm}
\epsfig{file=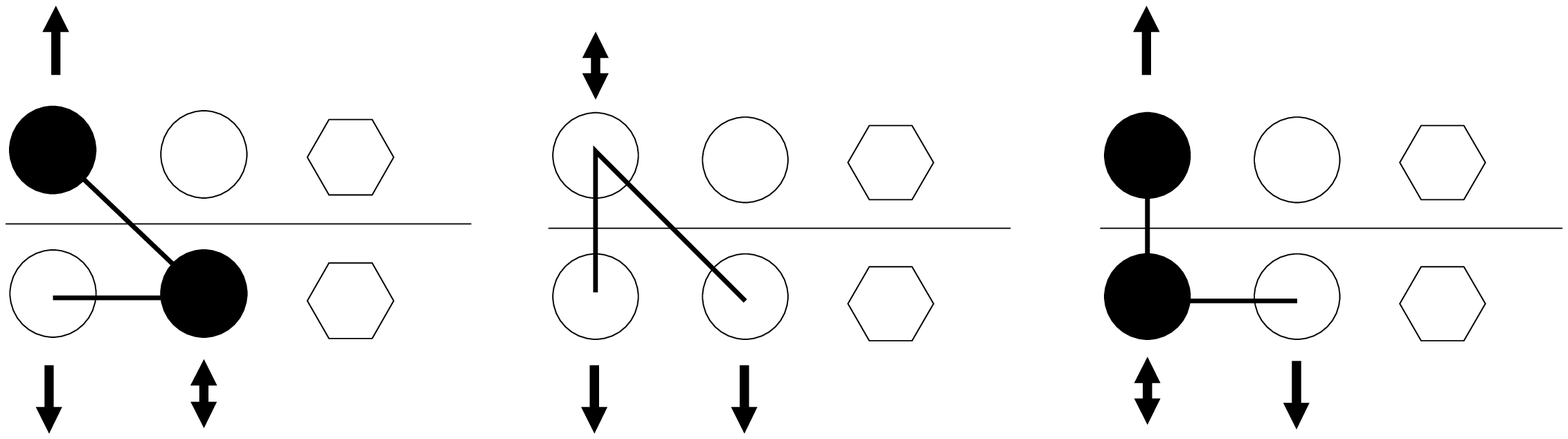,width=0.3\textwidth,angle=0}
\caption{ Example of a block of dimension $2$ (top) and $3$ (bottom) of
Eqs.\ (\ref{vectoreq}). (Notation see Fig. \ref{allplaqs}.) 
In- and possible out-going legs are indicated by
arrows and connected by a line. (Note that 
 in-going legs may always be out-going legs.) Upon flipping the two connected legs
and interchanging in- and out-going leg one diagram becomes another.
The other blocks of the same dimension may be obtained by reflection
 operations of the plaquettes.
}
\label{plaqloop1}\end{figure}

For a given plaquette --- associated with an operator $h^{(t)}$---
and a given in-going leg, whose site index is $n_i\in\{n,n+1,n+2\}$,
we can directly tell the dimension of the corresponding
block of $a$ and give a solution for its entries.
There are three cases to be considered:

I) For $t=1,2$ and $n_i=n+2$ (or $t=3,4$ and $n_i=n$)
we get only two equations from Eqs.\ (\ref{vectoreq}).
Since both  plaquette states (corresponding to non-diagonal $h^{(t)}$)
have equal weights, we can put the bounce weight to zero.
 The proceeding of the loop is then deterministic.
(see  top of Fig.\ \ref{plaqloop1}.)

II)  Otherwise  if
$(t=5,\,n_i\neq n+1)$ or $(t\neq 5,\,n_i=n+1)$
we have to consider three equations which may be treated as in the SSE.
(see Eqs.\ (\ref{soldim3}) and bottom of Fig.\ \ref{plaqloop1}.)

\begin{figure}
\epsfig{file=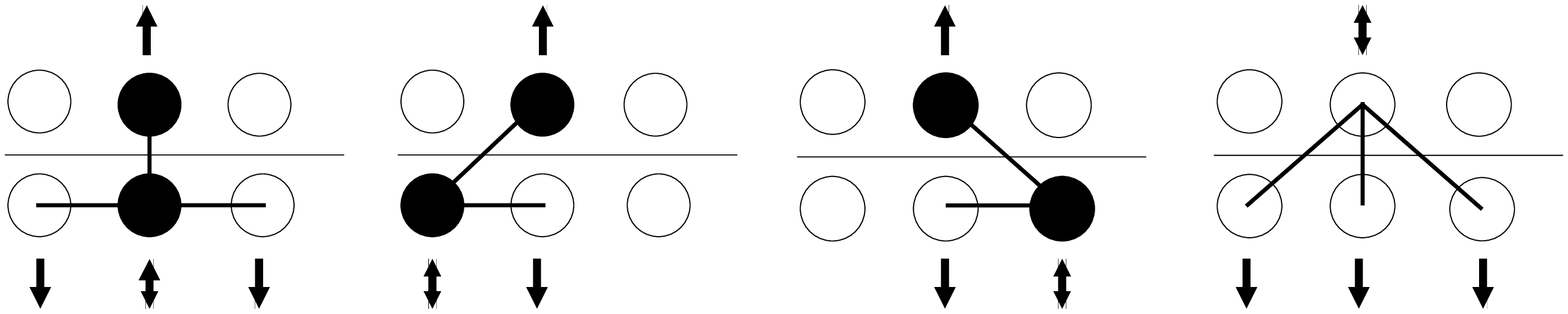,width=0.45\textwidth,angle=0}
\rule{\hsize}{2pt}

\vspace{0.2cm}
\epsfig{file=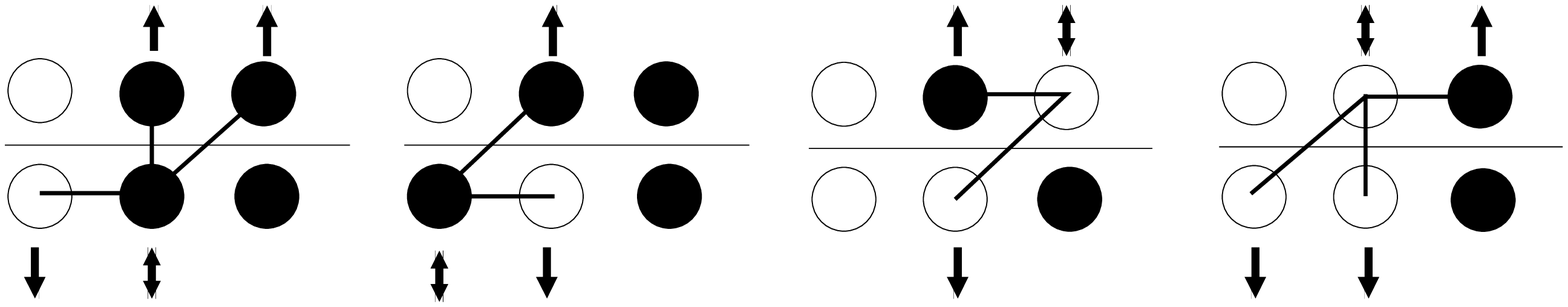,width=0.45\textwidth,angle=0}
\caption{ Two examples of a block of dimension $4$ of
Eqs.\ (\ref{vectoreq}) similar to Fig.\ \ref{plaqloop1}.
The other blocks of the same dimension may be obtained by reflection with respect to the
horizontal or vertical axis.
}
\label{plaqloop3}\end{figure}

III) 
In  the remaining cases we need to solve a sub-system of Eqs.\ (\ref{vectoreq})
of not less than four equations. (see Fig.\ \ref{plaqloop3}.)
Among the four  plaquette states involved in this block
we find two states corresponding to non-diagonal $h^{(t)}$
---which we call $i_n$ and $j_n$. From Fig.\ \ref{plaqloop3} we infer
that if $i_n$ corresponds to $t\in\{1,2\}$, $j_n$ corresponds to
$t\in\{3,4\}$, et vice versa.
The remaining two states correspond to diagonal operators; we call them
$i_d$ and $j_d$. The entries of $a$ are given by
\begin{eqnarray*}
a_{i_n,i_d}&=&\max\{[W(i_d)/2+W(i_n)-W(j_d)/2]/2,0\} \\
 a_{i_n,j_d}&=&\min\{[W(j_d)/2+W(i_n)-W(i_d)/2]/2,W(i_n)\}\\
 a_{i_d,j_d}&=&\min\{[W(i_d)+W(j_d)-2W(i_n)]/2,W(i_d)\}\\
a_{j_d,j_d}&=&\max\{0,W(j_d)-W(i_d)-2W(i_n)\}\\
 a_{j_n,k}&=&a_{i_n, k},\quad k=i_d,j_d.\\
\end{eqnarray*}
Note that we assumed $W(i_d)\leq W(j_d)$ and exploited $W(i_n)=W(j_n)$.
With this choice only one  diagonal entry 
(namely, $a_{j_d,j_d}$) may be non-negative.
To be concrete: only  
 the situation depicted in the upper part of Fig.\ \ref{plaqloop3}
admits $a_{j_d,j_d}=(J_z-J_x)/2>0$ if $J_z>J_x$.

The SCSE is
 more intricate (one has to consider several cases separately) and
 the bounce cannot always be
avoided, but there  is a considerable improvement 
(at low temperatures) with respect to the SSE.

\begin{figure}
\epsfig{file=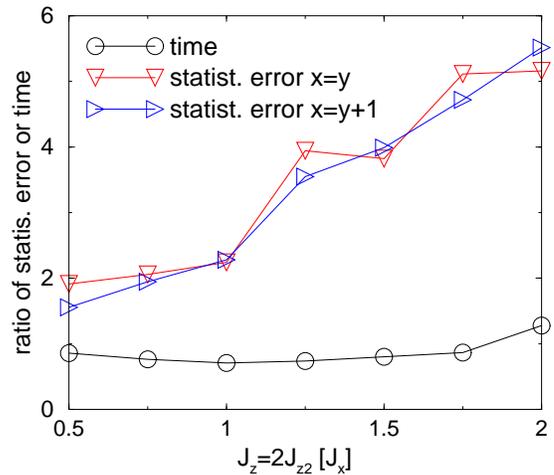,width=0.4\textwidth,angle=0}
\caption{We compare
simulations ($J_z=2J_{2z}$, 192 sites, $T=0.01J_x/k_B$, $2\cdot 10^{4}$
 MC-sweeps)
for SSE and SCSE.
Shown is the ratio of the time consume of the SSE and SCSE simulations
as well as the ratio of the mean (over the first 10
 Matsubara frequencies) statistical error of SSE and SCSE. 
$x=N/2$ and $y$ are defined in Eq.\ (\ref{intbypart2}).  }
\label{g2err}\end{figure} 

{\em Numerical results}---In this paper we use the method proposed in 
Ref.\ \onlinecite{techpap} to calculate the conductance for the Hamiltonian
 (\ref{hamilfrus}).  We evaluate the expression
 \begin{equation}\label{intbypart2}g(\omega_M)=-\omega_M/\hbar\int_0^{\hbar\beta} \cos(\omega_M
\tau)\langle P_xP_y(i\tau)\rangle d\tau
\end{equation}
at the Matsubara frequencies $\omega_M=2\pi
M(\beta\hbar)^{-1},\; M\in{\mathbb N}$. The conductance is obtained by
extrapolating $g(\omega)$ from the Matsubara frequencies to $\omega=0$.
The operators in Eq.\ (\ref{intbypart2}) are given by  
($e$ is the charge unit) $P_x=e\sum_{n>x}S^z_n$.
The $\omega=0$-value of $g$ does not depend on $x$ or $y$.\cite{techpap}

To illustrate the improvement gained by introducing SCSE
 we performed for a special choice of parameters ($J_z=2J_{z2}$)
simulations of $g(\omega_M)$.
In Fig. \ref{g2err} we plotted the ratios of the time consume
along with the ratio of the statistical error in $g(\omega_M)$
(for $M=1,\dots,10$).
The SSE is slightly faster (about 80 percent) but
 while the statistical error grows linearly with $J_z$
for SSE, it grows more modestly for the SCSE.

\begin{figure}[h]
\epsfig{file=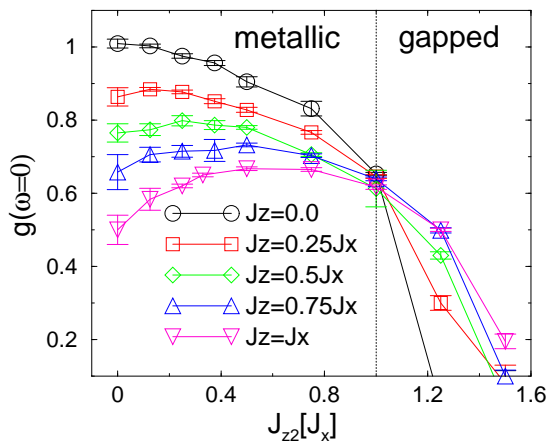,width=0.4\textwidth,angle=0}
\caption{Conductance of the Heisenberg chain (400 sites, $T=0.01J_x/k_B$)  with next
 nearest neighbor interaction $J_{z2}$. 
The phase boundary from Ref.\ \onlinecite{Kats} is displayed.
  (We use OBC's, $2\cdot 10^5$ MC-sweeps.)}
\label{gzzall}\end{figure}

{\em Frustrated system}---
Using the Lanczos-method Zhuravlev, Katsnelson and
coworkers\cite{Kats} obtained a very complete picture 
of the
 frustrated system given by the Hamiltonian  (\ref{hamilfrus}).
They set-up a low-temperature phase diagram with two gapped and a
gap-less
phase. In the gap-less phase the system is---for a large range of 
parameters---very well-described by the Luttinger liquid picture.\cite{Kats}
In Fig.\ \ref{gzzall} $g(\omega=0)$---extrapolated by a quadratic
fit---is plotted versus
$J_{z2}$ for various $J_z<J_x$.
For this region in parameter space
Ref. \onlinecite{Kats} finds a phase boundary between the gap-less
and the gapped phase at $J_{z2}=J_x$.

In a Luttinger liquid the conductance equals 
 the Luttinger parameter\cite{techpap,ApelRice,Giamarchi}
and may hence be viewed as an effective interaction.
Since $J_{z2}$ mediates a nearest neighbor attraction and thereby reduces the
 effective interaction,
  the conductance first grows with $J_{z2}$ and then
  assumes its maximum approximately when $2J_{z2}=J_z$ is satisfied.
We see that within error bars the conductance does not vary on the
phase boundary. 

As we work at a finite temperature of $k_BT=0.01J_x$, the conductance
goes smoothly to zero in the gapped phase such that a determination
of the phase boundary from Fig.\ \ref{gzzall} becomes difficult.


A remarkable fact of the phase diagram in Ref.\ \onlinecite{Kats}
is  the fact that the gap-less phase is not bounded in parameter space;
it contains, e.g.,  the line with $2J_{z2}=J_z$.
It is therefore interesting to study the behavior of the 
conductance on this line.
However, if we increase $J_z$ and
$J_{z2}$ simulating the conductance becomes more difficult. (The
statistical error grows.) 

\begin{figure}[h]
\epsfig{file=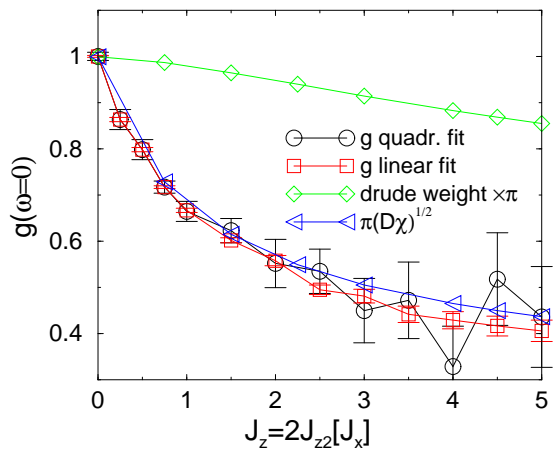,width=0.4\textwidth,angle=0}
\caption{Conductance of the Heisenberg chain (400 sites, $T=0.01J_x/k_B$)  with next
 nearest neighbor interaction $J_{z2}$ along the line in parameter space
 where $2J_{z2}=J_z$. Circles: quadratic extrapolation from the first three
 Matsubara frequencies, Squares: linear extrapolation from the first six
 Matsubara frequencies. (We use OBC's, $2\cdot 10^5$ MC-sweeps.) 
For comparison Exact diagonalization results are given.}
\label{metallic}\end{figure}

Only the SCSE allows us to compute
the conductance for as large interaction values as
 $J_z=5,\; J_{z2}=2.5$.
The conductance is plotted in Fig.\ \ref{metallic}.
Beside the statistical error we have an extrapolation error depending
on our extrapolation scheme. In Fig.\ \ref{metallic} we compared two
schemes: a quadratic fit to the first three
 Matsubara frequencies and a linear fit from the first six
 Matsubara frequencies. The former should give a smaller extrapolation
 error,
but it enhances the statistical error of $g(\omega)$.
The latter has a larger extrapolation
 error,
but it suppresses the statistical error of $g(\omega)$.
For small parameters the two fits almost coincide. But for larger
parameter values---when the statistical error increases---
the quadratic fit starts to fluctuate.

For this system  the DC-Drude weight $D$ and the susceptibility $\chi$
at $T=0$ may be obtained by exact diagonalization.\cite{Kats}
The same is true for the conductance via the relation
$g=\pi\sqrt{D\chi}$ valid for a Luttinger liquid.\cite{ApelRice,Schulz}
Fig.\ \ref{metallic} provides also the exact diagonalization data
showing good agreement with 
the SCSE data within error bars.

{\em Conclusion}---In this letter we showed that
 the performance of the SSE can be
substantially improved by selecting a different splitting
of the Hamiltonian. We used this strategy for an $xxz$-chain with
next-nearest-neighbor-interaction, but it should also apply 
to  systems which have additional plaquettes with a reduced number of
plaquette states.
These systems include fermionic chains with arbitrary interaction part,
 which may also be coupled by an interaction term as well as 
(not necessarily one-dimensional)
 spin systems that have at the same time Heisenberg-like and 
Ising-like interaction bonds.

\end{document}